\documentclass[12pt]{paper}
\usepackage{natbib}
\usepackage{subfig}
\usepackage[latin2]{inputenc}
\usepackage[T1]{fontenc}
\usepackage{graphicx}
\usepackage{amssymb}
\usepackage{bm}
\usepackage{amsmath}
\usepackage{color}
\begin{document}

\title{{Mind model seems necessary for the emergence of communication}}

\author{A. L{\H o}rincz\thanks{Corresponding author. \vspace{1mm}
\newline Address of all authors:
\newline \vspace{.2mm} Department of Information Systems
\newline \vspace{.2mm} E\"otv\"os Lor\'and University
\newline \vspace{.2mm} P\'azm\'any P\'eter s\'et\'any 1/C
\newline \vspace{.2mm} Budapest, Hungary H-1117 \vspace{1mm}
\newline Emails:
\newline \vspace{.2mm} A. L{\H o}rincz: andras.lorincz@elte.hu
\newline \vspace{.2mm} V. Gyenes: gyenesvi@inf.elte.hu
\newline \vspace{.2mm} M. Kiszlinger: kmelinda@inf.elte.hu
\newline \vspace{.2mm} I. Szita: szityu@gmail.com} \hspace{2mm}\& V. Gyenes \& M. Kiszlinger \& I. Szita}

\maketitle

\begin{abstract}
We consider communication when there is no agreement about symbols and meanings. We treat it within the framework of
reinforcement learning. We apply different reinforcement learning models in our studies and simplify the problem as
much as possible. We show that the modelling of the other agent is insufficient in the simplest possible case, unless
the intentions can also be modelled. The model of the agent and its intentions enable quick agreements about
symbol-meaning association. We show that when both agents assume an `intention model' about the other agent then the
symbol-meaning association process can be spoiled and symbol meaning association may become hard.
\end{abstract}

\section{Introduction}\label{s:intro}

The emergence of communication is one of the most enigmatic problems for several disciplines including evolution,
natural language theory, information technology. For a recent collection of papers, see, e.g.,
\cite{cangelosi05emergence}. For proper treatment, the concept of communication needs to be considered. First, let us
see a few examples.

\emph{Smoke signals.} These are `few bit' signals that could mean \verb"attention", \verb"danger", \verb"help", and so
on. The vocabulary is is small, the communication speed is high, the communication distance is large. The primary goal
of this communication is to overcome limited observation capabilities of other agents, to warn, and to coordinate
future actions.

\emph{Atomic interactions.} Not all light enabled interaction is, however, communication: Atoms, for example, interact
each other by exchanging photons. The emission and the absorption of photons are not intentional and the transmitted
photon has no hidden meaning.

\emph{Grooming.} According to Dunbar, grooming between monkeys is used, for example, to form alliances, serve, or
apologize \cite{dunbar97grooming}. Thus, we consider grooming communication, although it is non-verbal communication.

Then, the common features of communication are as follows: (i) communication is optional, (ii) it is intentional, and
(iii) communicated signals are symbols of certain meanings. Further, (iv) communication is successful, if the meaning
is the same for those who communicate. The emergence of communication is the subject of evolutionary linguistics (for a
recent review on evolutionary linguistics, see \cite{szamado06selective}). Evolutionary linguistics focuses on the
selective scenario that might give rise to the appearance of early languages. There are many theories and many
possibilities. Let us consider the popular and efficient language game approach
\cite{wittgenstein74philosophical,stenius67mood,steels97grounding,steels03trends}. In language games, the theoretical
approach makes certain assumptions. Presupposed conditions include the following: agents interact and their interaction
is `coordinated'. Thus, language game presupposes the existence of an agreement that agents start to engage themselves
in `coordinated actions'. Such an agreement is also a symbol-meaning association. Thus, the language game approach
assumes existing symbol-meaning association and builds on that assumption.

Our question concerns the very minimum of symbol-meaning association needed for successful communication. To this end,
we make the problem as simple as possible. Our analysis is embedded into the framework of reinforcement learning. We
study how communication may depend on the presence or the absence of the communication of emotions or internal
\emph{values}. In our simulations, communication will emerge as a deliberate action of the agents, but only if certain
conditions are fulfilled.

The paper is organized as follows. We provide the theoretical analysis in Section~\ref{s:theo}. This analysis shows the
necessity of emotional coupling between agents. We illustrate the analysis with simulations in simple scenarios
(Section~\ref{s:compstud}). We shall discuss our results in Section~\ref{s:disc}. We conclude in Section~\ref{s:conc}.
The paper is understandable without involved mathematical tools. Mathematical details are presented in the Appendices
for the sake of completeness.

\section{Theoretical analysis}\label{s:theo}

In this section, we investigate conditions when communication between two autonomous agents can \emph{emerge}. The
question is, how two autonomous agents could learn to communicate? We assume that neither the meanings nor the
communication signals are fixed in advance, there is no special method of negotiation, and there is no \emph{will} for
communication. However, the possibility for communication is given, and the world is such that communication could be
advantageous.

We investigate the problem in the framework of \emph{reinforcement learning} (for an excellent introduction, see
\cite{sutton98reinforcement}). We investigate how communication may emerge from \emph{joint} problem solving. That is,
we ask how agents could learn when and what to communicate based on utility; how they could learn to emit and interpret
signals provided that both parties benefit from those. It is surprising that if communication has a cost, then it is
still not sufficient that
\begin{itemize}
\item the possibility of communication is given and

\item communication would be beneficial for both agents (even with costs).
\end{itemize}

The underlying reason is that we have assumed that none of the agents has fixed an interpretation of the signals,
therefore they have to learn \emph{simultaneously} the translation from meanings to signals and vice versa. Let us call
one of the agents, that wishes to communicate something the `speaker', and the other one, which should learn to
interpret it, the `listener'. Now consider the case, when both agents are in the learning phase, and the speaker
experiments with different signals to express different meanings. The listener may not be able to differentiate the
meanings, and because of the costs, stops listening (i.e. learns that it is not worth to communicate). This effect
appears already in the simplest possible case. In this case, behaviors can be computed analytically.

Consider two agents, $A$ and $B$. For the sake of simplicity, we assume that communication is one-directional: $A$ may
speak and $B$ may listen to it. In each episode, agent $A$ may either be in state "1" or "2" (with equal probability),
and has three possible actions: \texttt{communicate "X"}, \texttt{communicate "Y"}, and \texttt{do not communicate}.
Communication has a cost of $1> c_A\ge 0$. Agent $B$ may listen to the signal of $A$ for a cost of $1> c_B \ge 0$, and
has to guess the state of $A$ (say \texttt{"1"} or \texttt{"2"}). They both receive a reward of $+1$, if the guess is
correct and a penalty of $-1$ if not. Since the cost of communication is less than the reward obtainable by it,
communication is desirable, if the two agents are able to agree that saying \texttt{"X"} means one of the states and
saying \texttt{"Y"} means the other.

\begin{figure}[h]
\begin{center}
   \includegraphics[width=6cm]{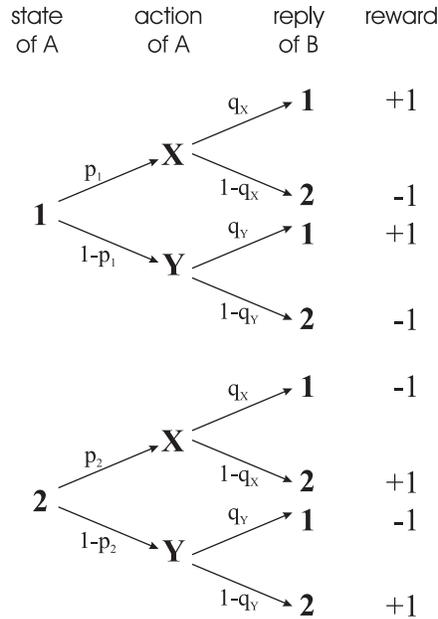}
   \caption{Various outcomes and associated rewards}\label{f:cases}
\end{center}
\end{figure}

The policy of $A$ can be described by the triple $M_A = (\alpha, p_1, p_2)$, where $\alpha$ is the probability that A
will communicate something, $p_1$ is the probability that $A$ says \texttt{"X"} in state \texttt{"1"}, $1-p_1$ is the
probability that $A$ says \texttt{"Y"} in state \texttt{"1"} given that he decides to communicate, and $p_2$ is the
probability that $A$ says \texttt{"X"} in state \texttt{"2"}, $1-p_2$ is the probability that $A$ says \texttt{"Y"} in
state \texttt{"2"} given that he is communicating. Similarly, the policy of $B$ can be described by the triple $M_B =
(\beta, q_X, q_Y)$, where $\beta$ is the probability that $B$ will listen to the signal, $q_X$ is the probability that
$B$ guesses \texttt{"1"} after hearing \texttt{"X"}, $1-q_X$ is the probability that $B$ guesses \texttt{"2"} after
hearing \texttt{"X"} given that he listens, and $q_Y$ is the probability that $B$ guesses \texttt{"1"} after hearing
\texttt{"Y"}, $1-q_Y$ is the probability that $B$ guesses "2" after hearing \texttt{"Y"} given that he listens. The
probabilities and rewards for the case when $A$ talks and $B$ listens are summarized in Figure~\ref{f:cases}. If $B$
does not listen, or $A$ does not talk, then $B$ guesses \texttt{"1"} with probability 0.5.


It is easy to calculate, that if both of them communicate, the common part of their expected reward is
$(p_1-p_2)(q_X-q_Y)$, and 0 if any of them is not communicating. Thus, the expected reward for policies $M_A$ and $M_B$
is
\[
  R_A(M_A, M_B) = \alpha \cdot (-c_A) + 2 \alpha\beta
  (p_1-p_2)(q_X-q_Y)
\]
for agent $A$ and
\[
  R_B(M_A, M_B) = \beta \cdot (-c_B) + 2 \alpha\beta (p_1-p_2)(q_X-q_Y).
\]
for agent $B$. We have assumed that neither $A$ nor $B$ can bind
meanings to signals, so initially $p_1 \simeq p_2$ and $q_X \simeq
q_Y$. Let us investigate the learning process of agent $A$. If
$|q_X - q_Y| < \varepsilon$ ($B$ cannot distinguish well between
meanings), the cost term of $A$ will be greater than his reward
term, so (i) he cannot tune $p_1$ and $p_2$ reliably (their
gradient is small), and (ii) he can minimize his losses by
lowering $\alpha$. The exact value of $\varepsilon$ depends on the
cost of communication. Similarly, $B$ will try to minimize $\beta$
until $A$ does not learn to distinguish between concepts, and
cannot reliably tune $q_X$ and $q_Y$.

As a result, during early trials, $p_1$, $p_2$, $q_X$ and $q_Y$ can only change stochastically, by random walk. As the
cost of communication grows, so does $\varepsilon$, and the time needed to exceed this limit by random walk grows
exponentially. However, during this time, $\alpha$ and $\beta$ keep diminishing. So by the time $A$ and $B$ could (by
chance) break the symmetry, and learn the distinction of meanings, they will learn that communication is not useful. We
note that in the general case, knowing the other \emph{agent's dynamics} (the parameter sets ($p_1$, $p_2$, $\alpha$)
and ($q_X$, $q_Y$, $\beta$)) does not always help; e.g., if the reward of one agent is not available to the other agent
and vice versa, or, if the rewards of the agents do not depend on each other's behaviors. In our two-state example
behaviors are coupled. Then, in theory, agents could use certain methods to estimate the hidden reward function of the
other agent. For example, non-direct implicit estimation is accomplished by the general policy gradient method: this
method -- up to some extent -- overcomes partial observations. It is so, because individual trajectories are considered
in this case. Successful estimation is, however, highly improbable in sophisticated real life situations.

\subsubsection*{Main hypothesis}

Within the framework of reinforcement learning, we have a single means to cure the flaw described previously; the
agents should be able to model each other's \emph{intentions}; the dynamics and the `goals' of the other agent. This is
possible if the values $R_A$ and/or $R_B$ are made available to them.

Now, the situation becomes different: agent $A$ can optimize $M_A$ for a fixed $M_B$. Although agent $A$ cannot modify
the policy of $B$, he can model, what would be rewarding for agent $B$. Furthermore, he considers the optimal
combination of the $M_A$ and $M_B$ strategies. Let us see the possible scenarios:

\subsubsection*{1-step modelling}

Optimizing $M_A$ for a fixed $M_B$ means calculating the conditional strategy
\[
  M_{A|B}(M_B) = \arg\max_{M_A} R_A(M_A, M_B),
\]
that is, $A$ can calculate, that if $B$ followed $M_B$, what would be the optimal choice for himself, i.e., for $A$.

\subsubsection*{2-step modelling}

If agent $A$ `knows' that he is using the conditional \emph{1-step modelling} strategy about agent $B$, then he might
as well suppose that $B$ does the same, i.e., agent $A$ might suppose that the strategy of agent $B$ is the following:
\[
  M_{B|A}(M_A) = \arg\max_{M_B} R_B(M_A, M_B),
\]
Now, agent $A$ can simply choose his optimal strategy:
\[
  M^*_A = \arg\max_{M_A} R_A( M_A, M_{B|A}(M_A) ).
\]

It might be worth noting that this abstract problem phrasing goes beyond the problem of communication; it is a general
learning problem. If an agent does something and it is visible to the other agent then it is a signal, which is
dependent on the state of the first agent. If both agents are learning, then the situation becomes similar to our
simplified example on communication.

Now, we introduce the basic concepts of reinforcement learning.
Then we turn to the illustrative numerical experiments.

\subsection{Basic concepts of reinforcement learning}

Reinforcement learning aims to solve behavior optimization based on immediate rewards. The main goal of optimization is
to maximize the long-term discounted and cumulated reward, the \emph{value}, or \emph{return} during the decision
making process. Reinforcement learning problems may be solved by value function estimation or by direct strategy
(policy) search methods. In \emph{value function estimation}, states or state-action pairs are assigned value estimates
that reflect the expected value of the long term cumulated and possibly discounted reward of choosing them. The agent
is not greedy and may not choose the optimal immediate reward, but it tries to act greedily according to this value
function: he selects the next state or action, which promises the optimal long-term (discounted and) cumulated reward
also called \emph{return}.

It is known that in partially observed environment, like in our case when the internal states of the agents may not be
observed, the \emph{direct policy search} method can be more efficient \citep{bertsekas96neuro-dynamic}. In this case,
the policy of the agent is explicitly represented in a parameterized form, and the parameters are updated so that the
described policy becomes optimal from the point of view of the \emph{return}. Policy gradient methods maximize the
expected \emph{return} by using gradient methods. The gradient of the \emph{return function} can be calculated
explicitly if the return function is known (see Appendix~\ref{s:explicit_polgrad}). However, general methods also exist
for cases when the reward function is not known explicitly (Appendix~\ref{s:general_polgrad}).

Here, we shall present numerical results for these methods. Note that in our simplified problem the immediate reward
and the long-term reward are identical. More sophisticated situations show the same phenomena \cite{gyenes06emotion}.

\section{Computer experiments}\label{s:compstud}

We have tested our theoretical analysis by conducting numerical experiments. We used policy gradient methods, and
various methods where the agents modelled each other. We studied the following cases:

\begin{description}
    \item[Method 1:] The agents did not model each other. In this case we studied value based methods and \emph{explicit policy gradient
    method}. We present results for \emph{explicit policy gradient method}.
    \item[Method 2:] The agents did not model each other directly, but use the \emph{general policy gradient method}.
    This method models the world and thus the other agent implicitly.
    \item[Method 3:] Agent $A$ estimated agent $B$'s dynamics, i.e. the parameters that determine $B$'s policy. In this case,
agent $A$ used a 1-step model of agent $B$. Thus, in this model, agent $A$ \emph{senses} the rewards of agent $B$ and
chooses the optimal policy accordingly. Agent $B$ did not model agent $A$ and applied the policy gradient method
    \item[Method 4:] Both agent $A$ and agent $B$ had access to the rewards of the other agent and estimated each other's dynamics.
    Both agents used a 1-step model of each other to choose their optimal policy
    \item[Method 5:] Both agent $A$ and $B$ had access to the rewards of the other agent and estimated each other's dynamics.
    Agent $A$ used a 2-step model of $B$, agent $B$ used a 1-step model of $A$ to choose an optimal policy
    \item[Method 6:] Both agent $A$ and $B$ had access to the rewards of the other agent and estimated each other's dynamics,
    and used a 2-step model of each other to choose their optimal policy
\end{description}
\begin{figure}[t!]
\begin{center}
   \includegraphics[width=12cm]{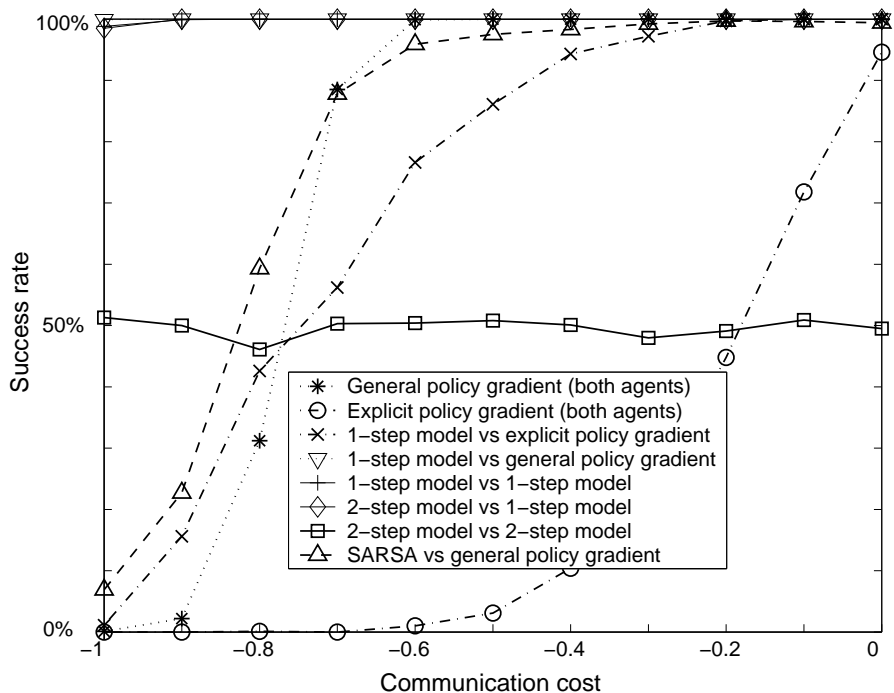}
   \caption{Performance of the various methods as a function of the cost of communication. Shorthand "vs": versus. For example,
   2-step vs 1-step: one agent used a 1-step model, the other agent used a 2-step model.}\label{f:success_rate}
\end{center}
\end{figure}
In the experiments, the values $\alpha$ and $\beta$ were initialized to $0.75$ in all cases. This choice enables us to
compare the different methods. The values are high enough to give a fair amount of chance for the agents at the
beginning to utilize communication. The values $p_1, p_2, q_X, q_Y$ were initialized randomly according to the uniform
distribution in the range $[0.4,0.6]$.

In the computational studies we averaged 1000 runs. In each run we had at most 1000 learning episodes. In each episode
an action was made by agent $A$ and a reaction, i.e., a guess, was made by agent $B$. Learning was considered
successful if after a certain number of steps, trials were 100\% successful; the reward in each of the next 100 trials
was +1. The number of steps needed for successful communication (not including the 100 successful ones that are used
for measuring success rate) is the time needed for the agreement. Figure~\ref{f:success_rate} depicts the success rate
for the different methods.

The general policy gradient method (Appendix~\ref{s:general_polgrad}) is superior to the explicit policy gradient
method (Appendix~\ref{s:explicit_polgrad}), however, if each agents uses these methods then they will not learn to
communicate if communication cost is high. Value estimation based reinforcement learning methods seem to be the weakest
amongst all methods that we studied (results are not shown here). Methods where agents use 1-step or 2-step models are
sometimes 100\% successful, with a single notable exception: if both agents use 2-step models then success rate is only
about 50\%. When rewards of the other agents are available then value estimation based method (the SARSA method
\cite{rummery95problem}) succeeds, too.

It can be seen, that when agents do not model each other, the chance that they learn to communicate decreases as the
cost of communication increases. However, when agents model each other, they are able to learn that communication is
useful even when the cost is high, with the peculiar exception when both agents use 2-step models.

\begin{figure}[h!]
\begin{center}
   \includegraphics[width=14cm]{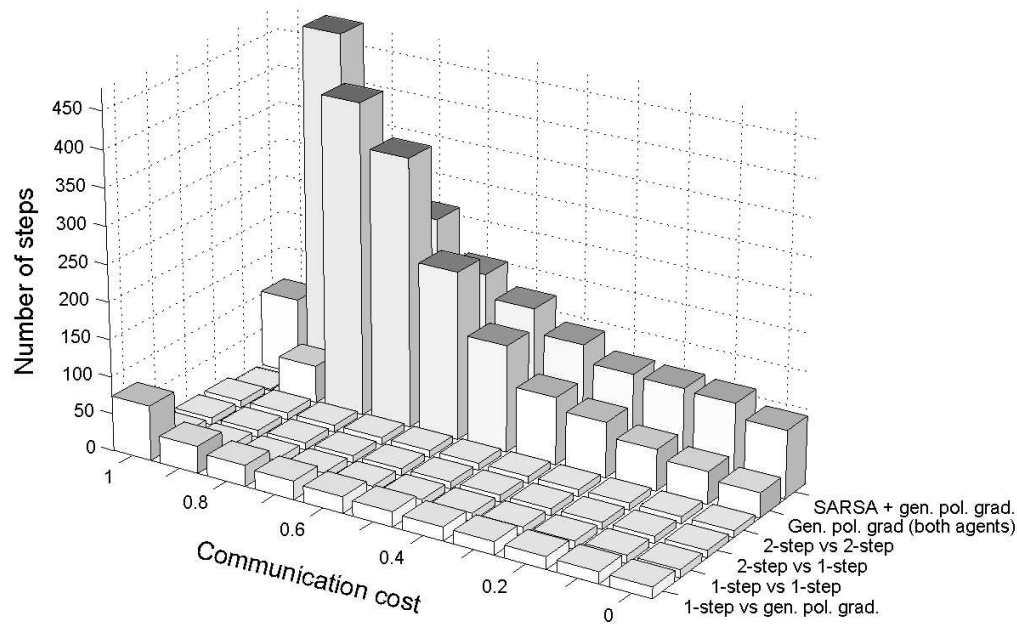}
   \caption{Learning time for various methods as a function of the cost of communication.
   Averages include only successful learning cases. Shorthand gen: general, pol: policy, grad: gradient}\label{f:success_time}
\end{center}
\end{figure}

Figure~\ref{f:success_time} depicts the time needed to reach an agreement. Situations when agreement was not reached
are excluded from these statistics.
It can also be seen, that when both agents can model the rewards of the other agent, then agreement about the
signal-meaning association is fast. This is so, because they shortcut the slow tuning procedure of reinforcement
learning. If this shortcut is not applied, like in the case of the value estimation based SARSA method, agreement can
be still reached, but only very slowly. When one of the agents thinks two steps ahead, agreement is even faster. In
this case, agreement is accomplished in 1 step after an initial transient of 10 steps when the agents estimate each
others' parameters. When both agents try to think two steps ahead and agreement is only achieved in 50\% of the cases,
agreement -- if it occurs -- is very fast. Thus, if agreement is not reached quickly, then agents could suspect that
the second-order intentional model (e.g., one agent assumes that the other agent uses a 1-step model) is not valid.

\section{Discussion}\label{s:disc}

Theory of reinforcement learning shows that globally optimal solutions can be learned `easily' under strict conditions.
The relevant condition for us is the Markov condition: information from the past does not help in improving decisions.
In other words, every information is encoded into the actual state of the agent and all state variables are amenable to
the agent for acting and learning. If this condition together with some other technical assumptions are fulfilled, then
the learning task is called Markov decision problem (MDP, see, e.g., \cite{sutton98reinforcement} and the references
therein).

The Markov condition is hardly met in real life. It is not met in our case either, because the parameters of decision
making of agent $A$ (or $B$) (i) are subject to experiences of agent $A$ (or $B$), i.e., they depend on the history,
(ii) these parameters are not available for agent $B$ (or $A$), and (iii) agent $B$ (or $A$) would benefit from knowing
these parameters. In this case the world is only partially observed and task is called partially observed Markov
decision problem (POMDP) (see, e.g., \cite{hauskrecht00value} and references therein).

This lack of information can be eased by modelling the other agent. The other agent might have many variables and a
large subset of those variables can be modelled by different means.  We demonstrated this by using policy gradient
methods. Both the explicit policy gradient and the general policy gradient method develop models of the `private'
parameters of the other agent: they model the state-action mapping, that is, the policy of the other agent. The
modelling process can be explicit: a particular model is assumed in this case, or implicit, when there is a general
parametrization in the policy gradient. Our simulations demonstrate that the performance of the explicit policy
gradient model is inferior to that of the general model. This observation can be traced back to the differences between
the methods: general policy gradient makes direct use of the immediate rewards, deals with individual state-action
sequences separately. Thus, the general policy gradient method -- up to some extent and indirectly -- takes into
account the intentions of the other agent. In the case of the model based explicit policy gradient method this
connection is highly remote: the same information enters the computation only after expected value computation. Value
estimation based methods (not shown here) have the same drawback and they are also inferior to the general policy
gradient method. These notes concern our simple scenario that does not fulfill the conditions of MDPs.

We have shown that the lack of a single quantity, the reward, makes a huge difference: not having access to the reward
of the other agent, the emergence of communication can be seriously limited if communication involves cost. The
assumption that communication is costly seems realistic, because communication takes time. Without access to the
rewards of the other agent, the higher the cost, the sooner the agents learn that communication is useless.

There are several exceptions to this simple observation. For example, if the policy of one of the agents is steady
(i.e., this agent is not learning), then this agent will act effectively as the teacher and the adaptive agent can
learn either the appropriate signal (if he is the speaker) or the appropriate meaning (if he is the listener).

The problem arises if the learning rates of the two agents are about the same. Then, to develop a successful
communication, they should be able to sense and then model (implicitly or explicitly) the immediate rewards, or the
cumulated rewards of the other agent. We shall call this capability emotional intelligence. It is satisfactory if one
of the agents has that capability. If an agent has emotional intelligence then the learning of symbol-meaning
association may become very efficient.

There are many ways to make this learning efficient, depending on what the agents assume about their partner. Consider,
for example, that both agents have emotional intelligence and both agents use this emotional intelligence when they
learn to communicate. Now, it makes a huge difference how they use the emotional information they have. For example,
indirect modelling of the situation occurs if we assume that the agents receive the same reward. Then we are in the MDP
domain and we can apply MDP methods such as SARSA \cite{rummery95problem}) -- without directly modelling the other
agent -- safely.

A large improvement was gained if both agents considered what is the best to them. Further, if (only) one of the agents
used that information to `anticipate' what the other agent might prefer to do in the next step, in reaction to his
action, then learning became even faster -- as it was expected from theory (Section~\ref{s:theo}).

However, learning is severely spoiled if both agents are clever enough and anticipate the next step of the other
agents. This has the following explanation: both agents suppose that the other is using a 1-step model to model him,
which, in this case, is false, because both agents use 2-step models. In this situation, in 50\% of the cases the
randomly generated initial parameters allow to reach an agreement just by chance. In the other 50\% no agreement is
reached.

As we have noted earlier, in this peculiar case the agents could suspect that the 1-step model they use about the other
agent is false: the other agent also considers `what is on his partner's mind'. Such consideration are the starting
points of game theory. However, the situation here can be different from game theory. In principle, our agents can
expect very fast agreement and they can become frustrated because of the lack of this quick agreement. Our agents are
also emotionally intelligent and they might sense the frustration of the other agent. That is, our agents might note
that their models are not valid and might come to a joint agreement. Thus, in our case, agents may use higher-order
intentional models and they will succeed.

An advantage of our formulation is that the agent might decide if he wants to optimize the sum of the two rewards
(cooperative agent), his own reward (selfish agent), the reward of the other agent no matter how much it costs
(altruistic agent), might decide to change this choice, and so on. These situations call for further investigations.

In our simple example, the immediate reward and the long-term reward were identical. Situations, where these two
quantities are different have also been studied \cite{gyenes06emotion}. The observations are about the same as in the
simple case that we presented here.

\section{Conclusions}\label{s:conc}

We have used explicit and implicit models in reinforcement learning. The world was partially observed, but otherwise it
was simplified as much as possible: we used two agents, two actions and two signals. We have shown that emotional
intelligence is necessary for the emergence of communications even in this simplest possible case. Numerical
simulations demonstrate that if the rewards of the other agent are available for modelling, then signal-meaning
associations can be learned quickly. The order of intentionality agents suppose in their models about the other agent
may give rise to problems, but the mere fact of the disagreement indicates that the models could be invalid. Novel
situations may arise: agents might decide about their attitude towards other agents.

\subsubsection*{Acknowledgments} We are grateful to G\'abor Szirtes for his comments on the manuscript. This material
is based upon work supported partially by the European Office of Aerospace Research and Development, Air Force Office
of Scientific Research, Air Force Research Laboratory, under Contract No. FA-073029. This research has also been
supported by an EC FET grant, the `New Ties project' under contract 003752. Any opinions, findings and conclusions or
recommendations expressed in this material are those of the author(s) and do not necessarily reflect the views of the
European Office of Aerospace Research and Development, Air Force Office of Scientific Research, Air Force Research
Laboratory, the EC, or other members of the EC New Ties project.

\section{Appendices: Algorithms and pseudo codes}\label{s:appendix}

\appendix

\section{Explicit policy gradient method} \label{s:explicit_polgrad}

In this case the explicit reward functions are available  for the two agents and they can calculate the gradients of
the parameter sets $M_A = (\alpha, p_1, p_2)$ and $M_B = (\beta, q_X, q_Y)$:
\[
  R_A(M_A, M_B) = \alpha \cdot (-c_A) + 2 \alpha\beta
  (p_1-p_2)(q_X-q_Y)
\]
for agent $A$ and
\[
  R_B(M_A, M_B) = \beta \cdot (-c_B) + 2 \alpha\beta (p_1-p_2)(q_X-q_Y).
\]
for agent $B$. As can be seen from the equations, each agent also needs to estimate the parameters of the other agent
in order to calculate its own expected reward.

\begin{table}[h]
 \hrule \vskip1pt \hrule \vskip2mm
\begin{tabbing}
xxx \= xxx \= xxx \= xxx \= \kill

\> $\varepsilon = 0.05, r, p \in [0,1]$ \\

\> for each test \\

\>\> $\alpha, \beta$ = 0.75, initialize $p_1, p_2, q_x, q_y$ to
random values \\

\>\> for each episode \textit{i = 1, ..., MAX\_EPISODES} do \\

\>\>\> \textbf{Agent A} \\

\>\>\> update the approximation of the parameters of $B$: $\hat
\beta, \hat q_{x}, \hat q_{y}$ \\

\>\>\> update own parameters by gradient: \\

\>\>\>\> $\Delta \alpha = -c_A + \hat \beta (r + p) + \hat \beta
(p_1-p_2)(\hat q_X-\hat q_Y) (r - p)$ \\

\>\>\>\> $\Delta p_{1} = \alpha\hat \beta (\hat q_X-\hat q_Y) (r -
p)$ \\

\>\>\>\> $\Delta p_{2} = -\alpha\hat \beta (\hat q_X-\hat q_Y) (r
- p)$ \\

\>\>\>\> $\alpha \leftarrow \alpha + \varepsilon \Delta \alpha$ \\

\>\>\>\> $p_{1} \leftarrow p_{1} + \varepsilon \Delta p_{1}$ \\

\>\>\>\> $p_{2} \leftarrow p_{2} + \varepsilon \Delta p_{2}$ \\

\>\>\> \textbf{Agent B} \\

\>\>\> update the approximation of the parameters of $A$: $\hat
\alpha, \hat p_1, \hat p_2$ \\

\>\>\> update own parameters by gradient: \\

\>\>\>\> $\Delta beta = -c_B + \hat \alpha (r + p) + \hat \alpha
(\hat p_1 - \hat p_2)(q_X - q_Y) (r - p)$ \\

\>\>\>\> $\Delta q_{x} = \hat \alpha \beta (\hat p_1 - \hat p_2)(r
- p)$ \\

\>\>\>\> $\Delta q_{y} = -\hat \alpha \beta (\hat p_1 - \hat
p_2)(r - p)$ \\

\>\>\>\> $\beta \leftarrow \beta + \varepsilon \Delta \beta$ \\

\>\>\>\> $q_{x} \leftarrow q_{x} + \varepsilon \Delta q_{x}$ \\

\>\>\>\> $q_{y} \leftarrow q_{y} + \varepsilon \Delta q_{y}$ \\

\>\> end for \\

\>end for

\end{tabbing}
 \vspace{-3mm}
 \hrule \vskip1pt \hrule \vskip1mm
 \caption{Pseudo-code of the explicit policy gradient method}
\end{table}

\section{General policy gradient method} \label{s:general_polgrad}

Let our policy $\pi$ depend on the parameters summarized in a vector $\theta \in \mathbb{R}^k$. Let $X$ be the set of
all possible trajectories in the task, and let $r(X)$ denote the reward collected in an episode. Then $\eta(\theta)$,
the value of the policy $\pi(\theta)$, is the expected value of the reward:

\[
\eta(\theta) = E\left[r(X)\right] = \sum_x r(x) q(\theta,x)
\]
where $E\left[.\right]$, denotes the expectation operator, $x \in X$ denotes a trajectory, $r(x)$ denotes the reward
collected while traversing trajectory $x$ and $q(\theta,x)$ is the probability of traversing trajectory $x$ having
parameters $\theta$. The gradient of $\eta(\theta)$ with respect to $\theta$ is:

\[
\nabla\eta(\theta) = \sum_x r(x) \nabla q(\theta,x) = \sum_x r(x) \frac{\nabla q(\theta,x)}{q(\theta,x)} q(\theta,x) =
E\left[ r(X) \frac{\nabla q(\theta,X)}{q(\theta,X)}\right]
\]
A sequence of trajectories $x^1, x^2,\dots, x^n$ give an unbiased estimate of $\nabla\eta(\theta)$:
\[
\hat{\nabla}\eta(\theta) = \frac{1}{N} \sum_{i=1}^{N} r(x^i) \frac{\nabla q(\theta,x^i)}{q(\theta,x^i)}
\]
Because of the law of large numbers: $\hat{\nabla\eta(\theta)} \rightarrow \nabla\eta(\theta)$ with probability 1. The
quantity $\frac{\nabla q(\theta,x)}{q(\theta,x)}$ is called likelihood ratio or score function.

Let the trajectory $x$ be a sequence of states $x_1, x_2, \dots, x_T$, and let $p_{x_t x_{t+1}}(\theta)$ be the
probability of moving from state $x_t$ to $x_{t+1}$ having parameters $\theta$. Then:

\[
\frac{\nabla q(\theta,x)}{q(\theta,x)} = \sum_{t=0}^{T-1} \frac{\nabla p_{x_t x_{t+1}}(\theta)}{p_{x_t
x_{t+1}}(\theta)} \ ,
\] which can be derived the following way:
\[q(\theta,x) = \prod_{t=0}^{T-1} p_{x_t x_{t+1}} \]
\[ \Rightarrow \log q(\theta,x) = \log \prod_{t=0}^{T-1} p_{x_t x_{t+1}}\ = \sum_{t=0}^{T-1} \log p_{x_t x_{t+1}}\]
\[ \Rightarrow \nabla \log q(\theta,x) = \sum_{t=0}^{T-1} \nabla \log p_{x_t
x_{t+1}} = \sum_{t=0}^{T-1} \frac{\nabla p_{x_t x_{t+1}}}{p_{x_t x_{t+1}}} \ ,
\] since $\nabla \log f(x) = \frac{\nabla
f(x)}{f(x)}$. This sum can also be accumulated iteratively.

\begin{table}[h]
 \hrule \vskip1pt \hrule \vskip2mm
\begin{tabbing}
xxx \= xxx \= xxx \= xxx \= \kill

\> $z_0 = \textbf{0} \in \mathbb{R}^k, \Delta_0 = \textbf{0} \in
\mathbb{R}^k$ \\

\> for each episode \textit{j = 1, ..., N} do \\

\>\> $R_0 = 0 \in \mathbb{R}$ \\

\>\> for each state transition $x_t \rightarrow x_{t+1}$ do \\

\>\>\> $ z_{t+1} = z_{t} + \frac{\nabla p_{X_t
X_{t+1}}(\theta)}{p_{X_t X_{t+1}}(\theta)}$ \\

\>\>\> $ R_{t+1} = R_t + \frac{1}{t+1} (r_t - R_t) $ \\

\>\> end for \\

\>\> $\Delta_{j+1} = \Delta_j + R_t z_t $ \\

\> end for \\

\> $\theta  \leftarrow \theta + \frac{\Delta_{N}}{N}$

\end{tabbing}
 \vspace{-3mm}
 \hrule \vskip1pt \hrule \vskip1mm
 \caption{Pseudo-code for the general policy gradient method} \label{t:1s_method}
\end{table}

In our case the algorithm is simplified, since each episode consists of one step (agent $A$ says something and agent
$B$ replies). Furthermore, we update the parameters after each episode, which means $N = 1$ in the above algorithm.
This way the two cycles boil down to one line of update after each episode:
\[ \theta  \leftarrow \theta + r \frac{\nabla
p_{s,a}(\theta)}{p_{s,a}(\theta)} \ . \] The respective gradients and probabilities can be calculated from the
parameters $\alpha, p_1, p_2, \beta, q_X, q_Y$:

\begin{table}[hbt]
\caption{Gradients and probabilities for agent $A$}
\begin{center} \label{t1}
\begin{tabular}{|c|c|c||c|c|c|}
  \hline
  state & action & $\nabla \alpha$ & $\nabla p_1$ & $\nabla p_2$ & probability \\
  \hline
  1 & X & $p_1$ & $\alpha$ & 0 & $\alpha p_1$ \\
  1 & Y & $1-p_1$ & $-\alpha$ & 0 & $\alpha (1-p_1)$ \\
  2 & X & $p_2$ & 0 & $\alpha$ & $\alpha p_2$ \\
  2 & Y & $1-p_2$ & 0 & $-\alpha$ & $\alpha (1-p_2)$ \\
  * & $\emptyset$ & -1 & 0 & 0 & $(1-\alpha)$ \\
  \hline
\end{tabular}
\end{center}
\end{table}

\begin{table}[hbt]
\caption{Gradients and probabilities for agent $B$}
\begin{center} \label{t1}
\begin{tabular}{|c|c|c||c|c|c|}
  \hline
  state & action & $\nabla \beta$ & $\nabla q_X$ & $\nabla q_Y$ & probability \\
  \hline
  X & 1 & $q_X$ & $\beta$ & 0 & $\beta q_X$ \\
  X & 2 & $1-q_X$ & $-\beta$ & 0 & $\beta (1-q_X)$ \\
  Y & 1 & $q_Y$ & 0 & $\beta$ & $\beta q_Y$ \\
  Y & 2 & $1-q_Y$ & 0 & $-\beta$ & $\beta (1-q_Y)$ \\
  * & $\emptyset$ & -1 & 0 & 0 & $(1-\beta)$ \\
  $\emptyset$ & 1 / 2 & 0.5 & 0 & 0 & $\beta$ \\
  $\emptyset$ & $\emptyset$ & -0.5 & 0 & 0 & $(1-\beta)$ \\
  \hline
\end{tabular}
\end{center}
\end{table}

In the tables the state or action denoted by $\emptyset$ means communicating nothing.

\section{1-step modelling}

In this case the agent calculates a conditional strategy that
optimizes $M_A$ and $M_B$ jointly, as discussed in the text.
Recall, that by joint optimization we mean that we can calculate
the conditional strategy
\[
  M_{A|B}(M_B) = \arg\max_{M_A} R_A(M_A, M_B),
\]
that is, $A$ can calculate, that if $B$ followed $M_B$, what would
the optimal choice of $A$ be. $A$ can estimate the parameters of
$B$ and thus can estimate his policy, $M_B = (\beta, q_X, q_Y)$.
The same is true vice versa, for agent $B$ estimating the policy
of $A$, $M_A = (\alpha, p_1, p_2)$. The parameters can be
estimated by the agents observing each other's behavior, and
approximating the parameters with their relative frequencies, that
is, the ratio of the occurrence frequencies of certain actions:

\begin{table}[h]
 \hrule \vskip1pt \hrule \vskip2mm
\begin{tabbing}
xxx \= \kill

\> $\hat \alpha = \frac{\text{\# episodes where A had chosen to
communicate}}{\text{\# all episodes so far}}$ \\ \rule{0pt}{18pt}%

\> $\hat p_1 = \frac{\text{\# episodes where A said "X" in state
"1" }}{\text{\# all episodes so far where A had chosen to
communicate}}$ \\ \rule{0pt}{18pt}%

\> $\hat p_2 = \frac{\text{\# episodes where A said "X" in state
"2" }}{\text{\# all episodes so far where A had chosen to
communicate}}$ \\ \rule{0pt}{18pt}%

\> $\hat \beta = \frac{\text{\# episodes where B had chosen to
listen}}{\text{\# all episodes so far}}$ \\ \rule{0pt}{18pt}%

\> $\hat q_X = \frac{\text{\# episodes where B guessed "1" after
hearing "X"}}{\text{\# all episodes so far where B had chosen to
listen}}$ \\ \rule{0pt}{18pt}%

\> $\hat q_Y = \frac{\text{\# episodes where B guessed "1" after
hearing "Y"}}{\text{\# all episodes so far where B had chosen to
listen}}$
\end{tabbing}
 \vspace{-3mm}
 \hrule \vskip1pt \hrule \vskip1mm
 \caption{Estimating parameters} \label{t:1s_method}
\end{table}

Then $M_{A|B}(M_B)$ can be derived analytically, and is the
following:

\begin{itemize}

\item if $B$'s will to use communication $(\hat \beta)$ is so low
that it is not worth using communication for $A$ because of his
own cost, then do not communicate anything,

\item otherwise, if $A$ is in state $1$, and $B$ is more likely to
answer $1$ to $X$ than to $Y$ $(\hat q_X > \hat q_Y)$, or if $A$
is in state $2$ and $B$ is more likely to answer $2$ to $X$ that
to $Y$ $(\hat q_X < \hat q_Y)$, then say $X$,

\item otherwise say $Y$

\end{itemize}

The conditional policy of agent B, $M_{B|A}(M_A)$, is essentially
the same, but using the estimated parameters of A $(\hat \alpha,
\hat p_1, \hat p_2)$.

\begin{table}[t!]
 \hrule \vskip1pt \hrule \vskip2mm
\begin{tabbing}
xxx \= xxx \= xxx \= \kill
 \> if $2\hat \beta < c_A$ \\
 \>\> do not communicate \\
 \> otherwise \\
 \>\> if ($A$ is in state $1$ and $\hat q_X > \hat q_Y$) or ($A$ is
 in state $2$ and $\hat q_X < \hat q_Y$) \\
 \>\>\> say \textit{X} \\
 \>\> otherwise \\
 \>\>\> say \textit{Y} \\
 \>\> end if \\
 \> end if
\end{tabbing}
 \vspace{-3mm}
 \hrule \vskip1pt \hrule \vskip1mm
 \caption{Pseudo-code of the 1-step modelling method for agent A} \label{t:1s_method}
\end{table}

\section{2-step modelling}

Supposing that $B$ uses 1-step modelling, A can think one step
further. Based on that, he can simply choose his optimal strategy:
\[
  M^*_A = \arg\max_{M_A} R_A( M_A, M_{B|A}(M_A) ).
\]
This optimal policy can also be derived analytically, and is the
following:

\begin{itemize}

\item if $B$'s will to use communication $(\hat \beta)$ is so low
that it is not worth using communication for $A$ because of his
own cost, or $A$'s will to use communication $(\hat \alpha)$ is so
low that it is not worth using communication for $B$ because of
his own cost, then do not communicate anything,

\item otherwise, if $A$ is in state $1$, and $\hat p_1 > \hat p_2$
(or if $A$ is in state $2$ and $\hat p_1 < \hat p_2$), then
suppose that $B$ traces this, and answers 1 (2) if $A$ says $X$,
so say $X$,

\item otherwise say $Y$

\end{itemize}

Again, the optimal policy for agent $B$ is essentially the same,
using the other's parameters.

\begin{table}[t!]
 \hrule \vskip1pt \hrule \vskip2mm
\begin{tabbing}
xxx \= xxx \= xxx \= \kill
 \> if $2\hat \beta < c_A$ or $2\hat \alpha < c_B$\\
 \>\> do not communicate \\
 \> otherwise \\
 \>\> if ($A$ is in state $1$ and $\hat p_1 > \hat p_2$) or ($A$ is
 in state $2$ and $\hat p_1 < \hat p_2$) \\
 \>\>\> say \textit{X} \\
 \>\> otherwise \\
 \>\>\> say \textit{Y} \\
 \>\> end if \\
 \> end if
\end{tabbing}
 \vspace{-3mm}
 \hrule \vskip1pt \hrule \vskip1mm
 \caption{Pseudo-code of the 2-step modelling method for agent A} \label{t:1s_method}
\end{table}

\section{SARSA}

The SARSA algorithm builds a table and computes the value of each entries. For the description of the algorithm, see,
e.g., \cite{rummery95problem,Singh00Convergence} and references therein.

\clearpage\newpage


\begin{thebibliography}{10}

\bibitem{cangelosi05emergence}
\emph{Special Issue on the Emergence of Language}, Connection Science
  \textbf{17} (2005), no.~3-4, 185--397, Editor: A. Cangelosi.

\bibitem{bertsekas96neuro-dynamic}
D.~P. Bertsekas and J.~N. Tsitsiklis, \emph{Neuro-dynamic programming}, Athena
  Scientific, Belmont, MA, 1996.

\bibitem{dunbar97grooming}
R.~Dunbar, \emph{Grooming, gossip, and the evolution of language}, Harvard
  University Press, Cambridge, MA, 1997.

\bibitem{gyenes06emotion}
V.~Gyenes, A.~Bontovics, and M.~Kiszlinger, \emph{Experimenting with emotional
  intelligence and echo-state networks}, Technical Report ELU-November-2006-1,
  E\"ot\"os Lor\'and University, Budapest, H, 2006, unpublished.

\bibitem{hauskrecht00value}
M.~Hauskrecht, \emph{Value-function approximations for partially observable
  markov decision processes}, Journal of Artificial Intelligence Research
  \textbf{13} (2000), 33--94.

\bibitem{rummery95problem}
G.~Rummery, \emph{Problem solving with reinforcement learning}, Ph.D. thesis,
  University of Cambridge, Cambridge, UK, 1995.

\bibitem{Singh00Convergence}
S.~Singh, T.~Jaakkola, M.~L. Littman, and Cs. Szepesv\'ari, \emph{Convergence
  results for single-step on-policy reinforcement-learning algorithms}, Machine
  Learning \textbf{38} (2000), 287--303.

\bibitem{steels03trends}
L.~Steels, \emph{Evolving grounded communication for robots}, Trends in
  Cognitive Sciences \textbf{7} (2003), 308--312.

\bibitem{steels97grounding}
L.~Steels and P.~Vogt, \emph{Grounding adaptive language games in robotic
  agents}, Proc. of European Conf. on Artificial Life (Cambridge Ma) (P.~Harvey
  and P.~Husbands, eds.), MIT Press, 1997,
  http://www.ling.ed.ac.uk/~paulv/ecal97.pdf.

\bibitem{stenius67mood}
E.~Stenius, \emph{Mood and language game}, Synthese \textbf{17} (1967),
  254--274.

\bibitem{sutton98reinforcement}
R.~S. Sutton and A.~G. Barto, \emph{Reinforcement learning: {A}n introduction},
  MIT Press, Cambridge, MA, 1998.

\bibitem{szamado06selective}
{Sz}. Sz\'amad\'o and E.~Szathm\'ary, \emph{Selective scenarios for the
  emergence of natural language}, Trends in Ecology and Evolution (2006), (in
  press).

\bibitem{wittgenstein74philosophical}
L.~Wittgenstein, \emph{Philosophical investigations}, Basil Blackwell, Oxford,
  UK, 1974, Translated by G. Anscombe.

\end{thebibliography}

\providecommand{\bysame}{\leavevmode\hbox to3em{\hrulefill}\thinspace}
\providecommand{\MR}{\relax\ifhmode\unskip\space\fi MR }
\providecommand{\MRhref}[2]{%
  \href{http://www.ams.org/mathscinet-getitem?mr=#1}{#2}
} \providecommand{\href}[2]{#2}

\end{document}